\documentclass[prb,showpacs,preprintnumbers,twocolumn,amsmath,amssymb,superscriptaddress]{revtex4}
\usepackage{graphicx}% Include figure files
\usepackage{color}
\usepackage{dcolumn}% Align table columns on decimal point
\usepackage{bm}% bold math
\usepackage{psfrag}
\def\bq{{\bf q}}

\newcommand{\be}{\begin{equation}}
\newcommand{\ee}{\end{equation}}
\newcommand{\bea}{\begin{eqnarray}}
\newcommand{\eea}{\end{eqnarray}}

\def\a{\alpha}
\def\b{\beta}
\def\e{\varepsilon}
\def\d{\delta}
\def\g{\gamma}

\def\o{\omega}

\def\s{\sigma}

\def\G{\Gamma}
\def\D{\Delta}
\def\O{\Omega}

\def\ra{\rightarrow}
\def\up{\uparrow}

\def\down{\downarrow}

\def\bk{{\bf k}}

\def\bq{{\bf q}}

\def\bQ{{\bf Q}}

\def\nn{\nonumber}
\def\lb{\label}

\def\pref#1{(\ref{#1})}

\newcount\bozza \bozza=0
\ifnum\bozza=1
\newdimen\shift \shift=-2truecm
\def\lb#1{%
{\label{#1}\rlap{\kern\shift{$\scriptstyle#1$}}}}
\else\def\lb#1{\label{#1}} \fi

\begin{document}
%@twocolumnfalse\endcsname
%\draft
\title{Nature and Raman signatures of the Higgs (amplitude) mode\\in the coexisting superconducting
  and charge-density-wave state} 

\author{T. Cea}
\affiliation{ISC-CNR and Dep. of Physics, ``Sapienza'' University of
  Rome, P.le A. Moro 5, 00185, Rome, Italy}
\email{tommaso_cea@libero.it,lara.benfatto@roma1.infn.it}
\author{L. Benfatto}
\affiliation{ISC-CNR and Dep. of Physics, ``Sapienza'' University of
  Rome, P.le A. Moro 5, 00185, Rome, Italy}
\date{\today}

\begin{abstract}
We investigate the behavior of the Higgs (amplitude) mode when
superconductivity emerges on a pre-existing charge-density-wave state. We
show that the weak overdamped square-root singularity of the amplitude fluctuations in
a standard BCS superconductor is converted in a sharp, undamped
power-law divergence in the coexisting state, reminiscent of the Higgs
behavior in Lorentz-invariant theories. This effect reflects in a
strong superconducting resonance in the Raman spectra, both for an
electronic and a phononic mechanism leading to the Raman visibility of the
Higgs. In the latter case our results are relevant to the interpretation of
the Raman spectra measured experimentally in NbSe$_2$.
\end{abstract}


\maketitle

\section{Introduction}

The emergence of collective excitations after spontaneous breaking of
a continuous symmetry is the mechanism at the heart of the mass
generation for scalar and vector bosons in the Standard
Model.\cite{weinberg,higgs_discovery} Its direct analogous in condensed matter physics
is the appearance of two collective modes as fluctuations of the
macroscopic (complex) order parameter in a superconducting (SC)
system.\cite{nagaosa} Indeed, amplitude fluctuations, which are energetically
costly, represent the analogous of the Higgs field, and phase
fluctuations, which are massless at long wavelengths, represent the
Goldstone mode that can be gauged away to make the electromagnetic
field massive, leading to the Meissner
effect. Interestingly, the first experimental
evidence of the Higgs particle\cite{higgs_discovery} occurred simultaneously to a renewed
interest in the literature on the behavior of the Higgs mode in
condensed-matter systems,\cite{varma_cm14} like e.g.superfluid cold
atoms\cite{podolsky_prb11,prokofev_prl12,podolsky_prl13,coldatoms1,coldatoms2},
where it can be excited by shaking the optical
lattice\cite{coldatoms1,coldatoms2}. 

On the other hand, in conventional BCS superconductors the Higgs mode
is usually very elusive, for two concomitant reasons. From one side,
in contrast to Lorentz-invariant bosonic theories, in BCS
superconductors the amplitude fluctuations do not identify a sharp
power-law resonance, but only a weak square-root singularity at twice
the SC gap $\Delta$,\cite{volkov73,kulik81} that is strongly
overdamped by quasiparticle excitations.  Thus, even in a fully-gapped
superconductor, the Higgs-mode spectral function is a broad feature
peaked exactly at the edge $2\Delta$ where also single-particle
excitations start to develop. In addition, in the weak-coupling BCS
limit the Higgs mode is expected to couple weakly to typical
spectrosocpic observables, like the current, probed by means of
optical measurement, and the charge, probed by means of Raman
spectroscopy.  Indeed, the Higgs mode is a scalar, so it does not
couple directly to the current, unless disorder breaks translational
invariance\cite{cea_prb14}, and it couples weakly to the charge, due
to the intrinsic particle-hole symmetry enforced by the BCS solution. Thus,
unless one strongly perturbs the system out of
equilibrium,\cite{shimano,carbone,shimano14} the Higgs resonance
remains hidden in conventional superconductors.

An exception to this rule seems to be the case of NbSe$_2$, a
low-temperature superconductor ($T_c=7$K) where superconductivity
emerges after a charge-density-wave (CDW) transition at higher
temperature $T_{CDW}=33$ K. In this material a strong resonance in the
Raman response\cite{klein_prl80,sacuto_prb14} appears below $T_c$ at
about $2\Delta$. As suggested long ago within a phenomenological model
by Littelewood and Varma,\cite{varma_prb82} and within a more
microscopic approach by Brouwne and Levin,\cite{brouwne_prb83} this
peak can be assigned to the Higgs mode. Indeed, according to this
interpretation the pre-existing CDW state provides a soft phonon,
Raman active already below $T_{CDW}$,\cite{tsang_prl76} that  below
$T_c$ couples also to the Higgs mode. As a consequence, the phonon response
itself carries out a signature of the SC amplitude mode, that
becomes sharp since it is pushed slightly below the treshold $2\Delta$ of quasiparticle
excitations. In other words, the current theoretical understanding is that the
pre-existing CDW state is crucial to provide a mechanism of Raman
visibility of the Higgs mode, but is does {\em not} change its nature.

This interpretation seems to be supported by the recent
observation\cite{sacuto_prb14} that no sharp peak appears in the
isostructural superconducting NbS$_2$, which lacks CDW order. On the
other hand, the overall strength of the Higgs resonance in the
experiments is much larger than what predicted by the previous
theoretical work,\cite{varma_prb82,brouwne_prb83} especially when one
takes into account residual damping effects, neglected so far. In
addition, the Higgs signature shows a still unexplained dependence on
the Raman light polarization,\cite{sacuto_prb14} that could be used
instead to solve the still on-going debate on the symmetry of the CDW
and SC gaps in this material.\cite{arpes1,arpes2,arpes3,arpes4,arpes5}

In the present paper we address the above issues by unveiling the real
character of the Higgs fluctuations in the mixed CDW-SC state. As a
first result we show that in a CDW superconductor the CDW contributes crucially to
modify the Higgs mode itself, making its detection easier whatever is
the mechanism making it physically observable. By computing the Higgs
mode in a microscopic CDW+SC model we show that the presence of a CDW
gap above the SC one pushes the quasiparticle continuum away from the
Higgs pole at $2\Delta$, transforming the weak (overdamped)
square-root singularity of a conventional BCS superconductor in a
well-defined power-law divergence. Thus, even in the weak-SC-coupling
limit, the amplitude mode in the CDW+SC state resembles closely the
one found in Lorentz-invariant relativistic theories, suitable in the
bosonic limit.\cite{podolsky_prb11,prokofev_prl12,podolsky_prl13} This
result has its own interest in the current discussion on the nature of
Higgs fluctuations in condensed-matter systems, and it could be
further tested by non-equilibrium spectroscopy, where a direct
non-linear coupling of the electromagnetic field to the amplitude mode
can be generated.\cite{shimano14} Second, we compute explicitly the
Raman response of the Higgs mode both in the absence and in the
presence of an intermediate CDW phonon. In the former case the
coupling of the amplitude fluctuations to the Raman response is not
zero when computed in a lattice mode, but still very small. While in an ordinary
superconductor this would still lead to a negligible visibility of the
strongly overdamped Higgs mode, in the CDW case the modified Higgs
spectral function will emerge naturally as a strong resonance. In the
case where a CDW phonon is also present the computation of the Raman
response requires to account for the intermediate electronic
processes that make the CDW phonon itself Raman visible below
$T_{CDW}$. This mechanism is analogous to the one discussed
e.g. for the Raman-active phonons in CDW dichalcogenides\cite{nagaosa_raman,klein_raman82}
or for the optical-active phonons in carbon-based compounds\cite{rice_76} and 
few-layers graphene.\cite{kuzmenko,li,cappelluti_prb12} When this effect is taken into
account one finds that the double-peak (phonon+Higgs) Raman
structure emerging below $T_c$ has in general a non-trivial 
temperature and polarization dependence. These results suggest that the recent
observations of a polarization dependence of the Higgs signatures in
the Raman spectra of NbSe$_2$\cite{sacuto_prb14} can be used to disentangle 
the underlying symmetry of the two CDW and SC order parameters in this material.

The structure of the paper is the following. In Section II we
introduce the microscopic SC+CDW model and we compute the Higgs spectral function in the mixed
state, showing its remarkable difference with the case of a
conventional superconductor. In Section III we analyze instead the
Raman response, in the case where the CDW has an electronic 
(Sec. IIIA) or phononic (Sec. IIIB) origin. In both cases we explain all
the microscopic mechanisms making the Higgs and/or the phonon Raman
visible, and we stress the effect of a modified Higgs mode on the Raman
response. The final remarks are discussed in Section
IV. In Appendix A we discuss the differences and analogies between the
coupling of the Higgs mode to the charge density and the Raman
density, respectively. Finally, Appendix B contains some results for
the present model away from half-filling, to show the robustness of
our conclusions in a regime where an analyitical interpretation of the
numerical results cannot be given. 

\section{Higgs mode in the coexisting CDW+SC state}

As discussed in the Introduction, to make progress with respect to previous
work\cite{varma_prb82,brouwne_prb83} we need two ingredients: (i) an
overlap bewteen the SC and CDW gaps and (ii) a
lattice model, crucial to account for the Raman-polarization
effects. To minimize the resulting  technical complications we
choose here a single-band model on the square lattice (lattice spacing
$a=1$)  with band
dispersion $\xi_\bk\equiv\e_\bk-\mu=-2t(\cos k_x+\cos k_y)-\mu$ ,
where $t=1$ is the hopping (that fixes the energetic units from now
on) and $\mu$ is the chemical potential. Near half filling ($\mu=0$)
the nesting of the Fermi surface at the CDW vector $\bQ=(\pi,\pi)$
allows for a CDW instability to occur, with new bands
$\xi_\pm=-\mu\mp\sqrt{\e_\bk^2+D_0^2\gamma_\bk^2}$.  Here we model it
with an order parameter $D_0=W\sum_{\bk\s} \langle \gamma_\bk
c^\dagger_{\bk\s} c_{\bk+\bQ\s}\rangle$, where the $\g_\bk=|\cos
k_x-\cos k_y|$ factor modulates the CDW in the momentum space, so that
below $T_{CDW}$ the Fermi surface consists of small pockets around
$(\pi/2,\pi/2)$. This feature is reminiscent of the situation in
NbSe$_2$, where ARPES
experiments\cite{arpes1,arpes2,arpes3,arpes4,arpes5} reported the
presence of ungapped Fermi arcs below $T_{CDW}$. Here the coupling $W$
can be thought to originates either from an electronic interaction or
from the coupling to a phonon, as we shall discuss in Sec. IIB below. The
superconductivity originates from a BCS-like interaction term $
H_{SC}=-(U/N)\sum_{q} \Phi_\D^\dagger(\bq)\Phi_\D(\bq)$, where
$\Phi_\D(\bq)\equiv\sum_\bk c_{-\bk+\bq/2\down}c_{\bk+\bq/2\up}$ is
the pairing operator and $N$ is the number of lattice sites. When treated at mean-field level it leads to the
following Green's function $G_0^{-1}(\mathbf{k},i\omega_n)$, defined
on the basis of a generalized 4-components Nambu spinor
$\Psi^\dagger_\bk(i\o_n)\equiv(c^\dagger_{\mathbf{k}\uparrow}(i\omega_n),
c^\dagger_{\mathbf{k}+\mathbf{Q}\uparrow}(i\omega_n),
c_{-\mathbf{k}\downarrow}(-i\omega_n),
c_{-\mathbf{k}-\mathbf{Q}\downarrow}(-i\omega_n))$ that accounts for
the CDW band folding:
\begin{equation}
\lb{G0}
G_0^{-1}(\mathbf{k},i\omega_n)\equiv
i\omega_n-
\begin{pmatrix} \hat h && -\Delta_0 \sigma_0\\
-\Delta_0 \sigma_0 && -\hat h
\end{pmatrix},
\end{equation}
where $\Delta_0$ is the SC gap, $\sigma_i$ denotes the Pauli matrices and $\hat h$ is a $2\times2$
matrix:
\be
\hat h=
\begin{pmatrix} \e_\bk-\mu && -D_0\gamma_\bk\\
 -D_0\gamma_\bk && -\e_\bk-\mu
\end{pmatrix}.
\ee
The eigenvalues of the matrix $\hat h$ represent the two CDW bands $\xi_\pm$,
while in the SC state the full Green's function \pref{G0} has four
possible poles, corresponding to the energies $\pm E_\pm(\bk)$, with
$E_\pm=\sqrt{\xi_\pm^2+\D_0^2}$. We note in passing that a similar
model (but with $d$-wave symmetry of the SC order parameter) has been
the subject of intense investigation in the past within the context of
cuprate superconductors, as a potential model for a CDW-like pseudogap
phase. \cite{nota_ddw}

\begin{figure}[htb]
%prop_a.eps
\includegraphics[width=8cm,clip=true]{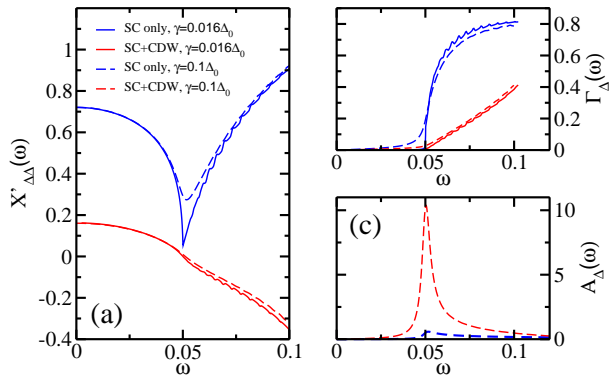}
\caption{Frequency dependence of the real (a) and imaginary (b) part
  of the Higgs spectral function (c) at $T=0$ in the presence (SC+CDW) and
  in the absence (SC only) of CDW, for a quasiparticle damping $\gamma=0.016\D_0$ (solid lines) and
$\gamma=0.1\D_0$ (dashed lines). Here $\D_0=0.025$, as in Fig.\
\ref{fig-scd}}. 
%In the pure SC state $X'_{\D\D}$ vanishes
%at $2\Delta$ with a square-root behavior, that is easily cut-off by a small
%residual quasiparticel scattering, while in the  CDW+SC $X'_{\D\D}$
%vanishes linearly at the Higgs mass. The imaginary part increases strongly above
%$2\Delta$ due to the activation of single-particle excitations: however, in the SC state
%the presence  of the CDW gap leads to a much smaller damping of the
%amplitude mode, making it possible the sharp resonance seen in Fig.\ \ref{fig-scd}.}
\label{fig-prop}
\end{figure}

To study the SC fluctuations we derive the effective action for the
collective modes by means of the usual Hubbard-Stratonovich decoupling of
$H_{SC}$.\cite{benfatto_prb04} After integration of the fermions one is
left with an action $S=S_{MF}+S_{FL}$ for the collective SC degrees of
freedom only, where $S_{MF}={N\Delta_0^2}/{TU}-\mathrm{Tr} \log G_0^{-1}$ is
the mean-field action and $S_{FL}[\Delta,\theta]$ is the action for the
amplitude $\Delta$ and phase $\theta$ fluctuations around the BCS solution
\pref{G0} above.  With respect to the usual case here the presence of a CDW
instability introduces in general in $S_{FL}$ a finite coupling between the
SC fluctuations at $\bq$ and $\bq+\bQ$, see Appendix B. However, one can
show that this coupling vanishes at half-filling, so for the sake of
simplicity we will discuss in the following this case. At the same time, in
the SC+CDW state the amplitude and phase/charge fluctuations remain
decoupled at Gaussian level, as it occurs in the usual weak-coupling SC
state,\cite{varma_prb82,benfatto_prb04} so we can neglect them in
what follows (see also discussion in Appendix A). The Gaussian action for
the amplitude SC fluctuations then reads
\be
\lb{ssc}
S_{FL}=\frac{1}{2}\sum_q \left(\frac{2}{U}+\chi_{\D\D}(q)\right)
  |\D(q)|^2,
\ee
where $\chi_{\D\D}$ is the response function for the amplitude operator
$\Phi_\D$ and $q=(\bq,\O_n)$, with $\O_n$ bosonic 
Matsubara frequencies.  The 
behavior of the amplitude fluctuations is controlled by the
function $X_{\D\D}\equiv2/U+\chi_{\D\D}$ in Eq.\ \pref{ssc}, since
$\langle |\D(q)|^2\rangle=1/X_{\D\D}$. In
analogy with the usual SC case\cite{kulik81} one can replace the term 
$2/U=2\sum_\bk \tanh(\beta E_\bk/2)/E_\bk$ by means of
the self-consistent equation for $\D_0$, in order to get:

\be
\lb{xdd}
X_{\D\D}=\frac{2}{U}+\chi_{\D\D}=\frac{2}{N}\sum_\bk \frac{(i\O_n)^2-4\D_0^2}
{E_\bk [(i\O_n)^2-4E_\bk^2]}\tanh({\beta E_\bk}/{2})
\ee
where the summation is restricted to the reduced Brillouin zone and
$E_\bk=E_\pm=\sqrt{\e_\bk^2+\D_0^2+D_0\g_\bk^2}$. 
%The vanishing of
%$X_{\D\D}$ identifies the mass of Higgs mode. 
In first approximation one can estimate the integral
\pref{xdd} at $T=0$ by assuming that both the density of states $N_0$ and the CDW
gap are constant (i.e. $\g_\bk=1$). In this case one can easily obtain
that for $\omega\lesssim 2\D_0$:
\be
\lb{xddapp}
X'_{\D\D}\simeq (4\D_0^2-\o^2) \frac{N_0 \mathrm{atan}(\D_0/D_0)}{\o
  \sqrt{4D_0^2+4\D_0^2-\o^2}}.
\ee
In the pure SC case ($D_0=0$) the denominator diverges as a
square-root at $\o=2\D_0$,\cite{kulik81} signaling the proliferation of
single-particle excitations above this threshold, and $X'_{\D\D}\sim
\sqrt{4\D_0^2-\o^2}$ near the Higgs mass $\o=2\D_0$. In contrast, in
the mixed state the CDW gap pushes the quasiparticle continuum away
from the Higgs mode, leading to a finite denominator in Eq.\
\pref{xddapp}. This implies a {\em linear} vanishing of $X'_{\D\D}$
at $2\D_0$ and suppresses its imaginary part $\G_{\D}\equiv
-X"_{\D\D}$. These simple estimates are confirmed by the numerical
calculation done for the real band dispersion, as shown in Fig.\
\ref{fig-prop}a,b, where we report both $X'_{\D\D}$ and $\G_{\D}$ in the
two cases, with and without preformed CDW state. To account also for
residual quasiparticle scattering we added a finite constant
broadening $\gamma$ in the response functions. As one can see in
Fig.\ \ref{fig-prop}a, in the SC+CDW state $X'_{\D\D}$ vanishes
approximately linearly, in contrast to the square-root of the usual SC
case\cite{kulik81} computed with the same $\D_0$, and $\G_\D$ (see Fig.\ \ref{fig-prop}b) develops
smoothly above $2\D_0$, since only few quasiparticle excitations in the
regions with small CDW gap contribute to it. These two concomitant effects lead to a
dramatic sharpening of the Higgs spectral function
$A_\D(\o)=(1/\pi)\mathrm{Im}X_{\D\D}^{-1}(i\O_n\ra \o+iO^+)$ 
\be
\lb{spectral}
A_\D\equiv \frac{1}{\pi}
\frac{ \Gamma_\D(\o)}{X_{\D\D}^{'2}(\o)+\Gamma_\D^2(\o)},
\ee
shown in Fig.\ \ref{fig-prop}c for the case $\gamma=0.1\Delta$. 
While in the ordinary SC state we recover the typical overdamped
structure of the amplitude fluctuations, in the CDW+SC state the Higgs
mode becomes a well-defined excitation, rather insensitive to residual
quasiparticle excitations, that are pushed away from the Higgs pole by
the CDW gap. As we show in Appendix B, the same qualitative
features survive also at finite doping, even if part of the Fermi
surface remains ungapped below $T_{CDW}$.
The strong modifications of the Higgs spectral function shown in Fig.\
\ref{fig-prop}c, and due to the emergence of superconductivity on the
pre-existing CDW state, represent the first crucial result of our
work. Indeed, they imply that whatever is the mechanism that couple
the Higgs mode to a physical observable, its detection in the mixed state
becomes easier since the mode itself is much sharper than in an
ordinary superconductor. As an example of this mechanism we will
discuss in the next Section the case of the Raman response.

\section{Raman signatures of the Higgs mode}

\subsection{Electronic mechanism}

We first compute the Raman spectra in the case where the CDW has
an electronic origin, so that the Raman visibility of the
Higgs can only be due to a direct coupling between the SC amplitude
fluctuations and the Raman charge fluctuations. Within the
effective-action formalism the Raman response can be 
computed by introducing in the fermionic model a source term $\rho_R$  coupled to
the Raman density operator  $\Phi_R(\mathbf{q})\equiv
\sum_{\mathbf{k}\sigma}\Gamma(\mathbf{k})c^\dagger_{\mathbf{k}-\mathbf{q}/2,\sigma}c_{\mathbf{k}+\mathbf{q}/2,\sigma}$. 
Here $\Gamma(\mathbf{k})$ is the Raman vertex, determined by the
polarization of the incident and scattered photon\cite{review}. After integration of the
fermions $\rho_R$  appears as an additional bosonic field in the
action $S_{FL}$, that now reads
\bea
\lb{sapprox}
S_{FL}&=&\frac{1}{2}\sum_q\left\{\left(\frac{2}{U}+\chi_{\D\D}(q)\right) |\D(q)|^2+\right.\nn\\
&+&\left.2\rho_R(q)\D(-q)\chi_{R\D}(q)+|\rho_R(q)|^2\chi^0_{RR}(q) \right\}.
\eea
Here $\chi_{AB}(q)$ denotes the response functions computed with the $A,B$
operators, where $R$ identifies the Raman density $\Phi_R$, $\D$ the
pairing operator $\Phi_\D$, and $\chi_{RR}^0$ represents the BCS Raman
response function in the absence of the collective modes. Notice that
even if one included explicitly the Coulomb forces the Raman response
here would be unaffected, since phase fluctuations do not couple to the Raman
response at $\bq=0$, while charge fluctuations, that in general screen
$\chi^0_{RR}$ in the symmetric $A_{1g}$ channel, \cite{review} are
ineffective in the present model since the charge-Raman coupling
vanishes exactly at half-filling because of particle-hole symmetry.
Finally, the Raman
response is obtained by analytical continuation $i\O_n\ra \o+i\delta$  as 
$S_R=-\frac{1}{\pi}[1+n(\o)]\chi_{RR}''(\bq=0,\o)$, where $n(\o)$ the
Bose-Einstein distribution and   $\chi_{RR}$ is the Raman
susceptibility, computed from Eq.\ \pref{sapprox} as:
\begin{equation}\lb{chirr}
\chi_{RR}(q)=\left[\frac{\delta^2 {S}_{FL}}{\delta \rho_R(-q)\delta \rho_R(q)}\right]_{\rho_R=0}\quad.
\end{equation}
\begin{figure}[tb]
%a1g.eps
\includegraphics[width=8cm,clip=]{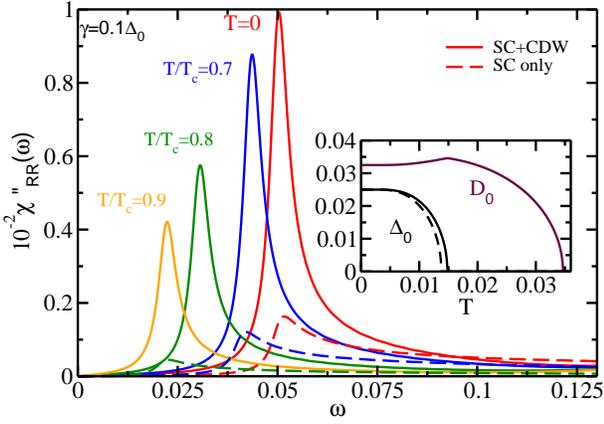}
\caption{ Electronic Raman response in the $A_{1g}$ channel in the
  coexisting (solid lines) and pure (dashed lines) 
SC state as a function of temperature. Inset: temperature evolution
of the SC gaps $\D_0$ and of the CDW gap $D_0$.}
\label{fig-scd}
\end{figure}

\begin{figure}[htb]
%diagrams.eps
\includegraphics[width=7.5cm,clip=]{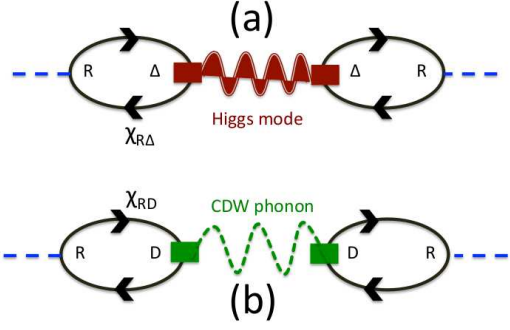}
\caption{Feynman diagrams representing the electronic processes
  responsible for the Raman visibility of (a) the Higgs mode when
  $g=0$ and (b) the
  CDW phonon above $T_c$. Here the dashed lines represent the incoming
  electromagnetic radiation, the full lines the electronic Green's
  functions $G_0$, the red and green wavy lines the Higgs
  propagator $X^{-1}_{\D\D}(\o)$ and the phonon one $D(\o)$, 
  respectively. In the coexistence state for $g\neq 0$ the two processes become
  interconnected, since both the phonon propagator (see Eq.\ \pref{om0new})
  and the $\chi_{RD}$ (see Eq.\ \pref{raman_full}) 
  are renormalized by the coupling bewteen the Higgs fluctuations and
  the CDW amplitude fluctuations.}
\label{fig-diagrams}
\end{figure}

By integrating out explicitly the amplitude fluctuations in Eq.\
\pref{sapprox} we are then left with the Raman response:
\be
\lb{chifull}
\chi_{RR}=\chi_{RR}^0-
{\chi^2_{R\D}}/X_{\D\D},
\ee
where the second term, depicted in Fig.\ \ref{fig-diagrams}a, is equivalent to compute the RPA
vertex corrections due to amplitude fluctuations.\cite{review} 
The coupling of the Higgs to the Raman density is mediated by the fermionic
susceptibility $\chi_{R\Delta}$:
\be
\lb{chird}
\chi_{R\D}=\frac{8\D_0}{N}\sum_\bk
\frac{\G(\bk)\e_\bk \tanh(\b E_\bk/2)}{E_\bk[(i\O_n)^2-4E_\bk^2]},
\ee
that is finite in the $A_{1g}$ channel where $\G(\bk)=(\cos k_x+\cos
k_y)\propto \e_\bk$. Since in the $A_{1g}$ channel the BCS contribution $\chi^0_{RR}$ is
negligible, for this symmetry the Raman response probes essentially the spectral
function $A_\D$ of the Higgs mode given by Eq.\ \pref{spectral}:
\be
\lb{chirrapp}
\chi''_{RR}\simeq\pi\chi^{'2}_{R\D}(\o)A_\D(\o),
\ee
leading to the results of Fig.\ \ref{fig-scd} for the CDW+SC (solid
lines) and SC only (dashed lines) case. As we discuss in detail in the
Appendix A, the coupling $\chi_{R\D}$ of the Raman response to the
Higgs is in general different from the coupling of the real charge
density to SC amplitude fluctuations, that vanishes at weak coupling
due to the approximate particle-hole symmetry of the BCS
solution.\cite{varma_prb82,benfatto_prb04} On the other hand, in
a ordinary SC state even if this is a finite quantity it multiplies a strongly
overdamped Higgs spectral function (see dashed lines in Fig.\ \ref{fig-prop}c), so the
overall Raman signature of the Higgs is the broad and weak feature
represented by dashed lines in Fig.\ \ref{fig-scd}. This justifies
on microscopic grounds the general expectation\cite{varma_cm14} that the
Higgs mode is irrelevant in the Raman
spectra of an ordinary superconductor. However, in a CDW+SC state the
situation is radically different, since the spectral function itself
of the Higgs mode is sharper than usual. Thus, even if in the
coexisting state the $\chi_{R\D}$ prefactor in Eq.\ \pref{chirrapp} is
smaller than in the usual SC state (see Appendix B), the modifications
of the Higgs spectral function $A_\D(\o)$ shown in Fig.\
\ref{fig-prop} lead to the sharp resonance at $2\Delta_0(T)$ shown in
Fig.\ \ref{fig-scd}. In other words, the CDW state is not modifying
the mechanism coupling the Higgs to the Raman probe, but it is
changing dramatically the nature of the Higgs mode itsefl, making it
detectable.

 \subsection{Phononic mechanims}

\begin{figure}[htb]
%a_phonon.eps
\includegraphics[width=8.5cm,clip=]{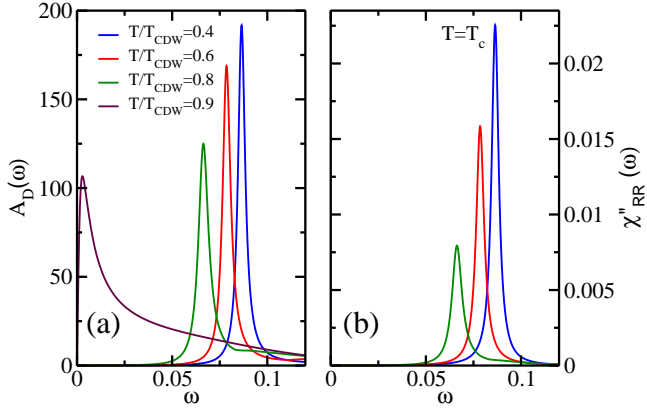}
\caption{Temperature evolution of the phonon spectral function (a) and
  of the Raman response (b) of the CDW phonon between $T_{CDW}$ and
  $T_c=0.4T_{CDW}$.}
\label{fig-phonon}
\end{figure}

A second possible mechanism that makes the Higgs mode visible in Raman
is the one proposed long ago in the case of NbSe$_2$,\cite{varma_prb82,brouwne_prb83} where
Higgs fluctuations come along with the CDW phonon. To explore also this possibility
we then consider the case where the CDW instability is driven
by the microscopic coupling of the electrons to a phonon of energy
$\omega_0$, $H_I=g\sum_{\bk\sigma} \g_\bk
c^\dagger_{\bk+\bQ\s}c_{\bk\s}(b^+_\bQ+b_{-\bQ})$, so that
$4g^2/\o_0$ is equivalent to the coupling $W$ used so far.  At
mean-field level the description of the CDW+SC state is then identical
to the one of Sec. II, and the SC amplitude fluctuations are described
by the same effective action \pref{ssc} derived above, leading to the
results of Fig.\ \ref{fig-prop}. However, in the calculation of the
Raman spectra we should account for two additional effects: (i) the
Raman response of the CDW phonon itself\cite{nagaosa_raman,klein_raman82} and (ii) the coupling between
the phonon and the Higgs, that reflects in a renormalization of the
phonon propagator.\cite{varma_prb82,brouwne_prb83} Since all
these issues have been discussed only in part in the previous
literature, we will outline the basic steps leading to the calculation
of the Raman response in this case. 

Let us first focus on the regime below $T_{CDW}$ but above $T_c$. When
the CDW is driven by a lattice instability the energy $\o_0$ of the
phonon is renormalized to the value $\O_0$ by the amplitude fluctuations
of the CDW.\cite{rice_ssc74} The renormalized phonon propagator below $T_{CDW}$
is then $D^{-1}(i\O_n)=-(\O_n^2+\O_0^2)/2\o_0$ with
\bea
\O_0^2&=&\o_0^2\left[1+\frac{2g^2}{\o_0}\chi_{DD}\right]=\nn\\
\lb{om0}
&=&\frac{4g^2\o_0}{N}\sum_\bk
\g_\bk^2\frac{(i\O_n)^2-4D_0^2\g_\bk^2}{E_\bk((i\O_n)^2-4E_\bk^2)}\tanh(\beta
E_\bk/2), 
\eea
where, in accordance to our notation, the subscript $D$ labels the CDW
amplitude operator $\Phi_D=\sum_{\bk\s}
\g_\bk(c^\dagger_{\bk\s}c_{\bk+\bQ\s}+h.c.)$. The second line of Eq.\
\pref{om0} has been derived by using the self-consistency equation for the
CDW gap, i.e. $1=(4g^2/\o_0N)\sum_\bk [\g_\bk^2 \tanh(\beta
E_\bk/2)]/E_\bk$. By comparison with Eq.\ \pref{xdd} one sees that $\O_0^2$
scales as the inverse propagator of the CDW amplitude fluctuations, that is
itself massless at about $4D_0$ (where the additional factor of two comes
from the $\gamma_\bk$ modulation factor).  As a consequence, the
renormalized phonon energy also follows the $T$ dependence of the CDW order
parameter, so it vanishes at $T=T_{CDW}$ and it is reduced with
respect to $\o_0$ at $T=0$,\cite{rice_ssc74} as shown by the phonon
spectral function $A_D(\o)=-(1/\pi)\mathrm {Im} D(i\O_n\ra \o+i\d)$
reported in Fig.\ \ref{fig-phonon}a.  Notice that, in contrast to previous
work,\cite{rice_ssc74,brouwne_prb83} we retained here the full frequency
dependence of $\O_0$, crucial as $T$ increases and $D_0\ra 0$. Thus, the
pole of the phonon propagator is determined self-consistently after
analytical continuation as as a solution of the equation
$\o^2-\O_0^2(\o)=0$.

The formation of the CDW state is
also responsible for the Raman visbility of the phonon. Indeed, in
analogy with what suggested for CDW
dichalcogenides\cite{nagaosa_raman,klein_raman82}, the
particle-hole excitations at $\bq=0$ probed by Raman couple to the
electronic CDW fluctuations at $\bQ$, that in turn can decay in a
phonon, as depicted by the process of Fig.\ \ref{fig-diagrams}b.  The
Raman response of the phonon above $T_c$ is then given by
\be
\lb{rrphonon}
\chi_{RR}(i\O_n)=-g^2\chi_{RD}^2D(i\O_n)=-\frac{2g^2\o_0\chi^2_{RD}}
{\O_0^2-(i\O_n)^2},
\ee
where
\be
\lb{chirdd}
\chi_{R D}={8 D_0}\sum_\bk
\frac{\G(\bk)\g_\bk^2\e_\bk}
{E_\bk((i\O_n)^2-4E_\bk^2)}\tanh(\beta
E_\bk/2),
\ee
In full analogy with the case of the $\chi_{R\D}$ function that makes the
Higgs mode Raman visible, the $\chi_{RD}$ susceptibility depends on the
combined symmetry of the Raman polarization, controlled by $\Gamma(\bk)$,
the lattice structure and the CDW symmetry. In the present case one can
easily see that Eq.\ \pref{chirdd} is different from zero only in the
$A_{1g}$ channel, where $\G(\bk)\propto \e_\bk$. 
Since $\chi_{RD}$ scales as the CDW order parameter $D_0$, the overall
temperature evolution of the Raman response \pref{rrphonon}  of
the phonon, shown in Fig.\ \ref{fig-phonon}b,  differs
considerably from the one of the spectral function. Indeed, as $T\ra
T_{CDW}$ the $\chi_{RD}$ vanishes rapidly (see inset of Fig.\
\ref{fig-full}a) and the Raman response is completely suppressed, in
agreement with the experimental observation in
NbSe$_2$.\cite{tsang_prl76}  It is also worth noting that the crucial role
played by the intermediate particle-hole excitations to control the
spectroscopic visibility of a phonon is a well-known effect in the
literature. For example, in the context of optical conductivity this is the
so-called charged-phonon effect, originally introduced by Rice for 
carbon-based compounds\cite{rice_76}, and widely discussed in the last few
years within the context of few-layers
graphene.\cite{kuzmenko,li,cappelluti_prb12} In this case it has been shown
that the strong doping dependence of the phonon-peak intensity and its
Fano-like shape\cite{kuzmenko,li} can be explained by computing the optical visibility of
the phonon coming from the process analogous to the one depicted in
Fig.\ \ref{fig-diagrams}, with the Raman vertex replaced in this case by
the current vertex.\cite{cappelluti_prb12}

\begin{figure}[htb]
%phonon_new.eps
\includegraphics[width=8cm,clip=true]{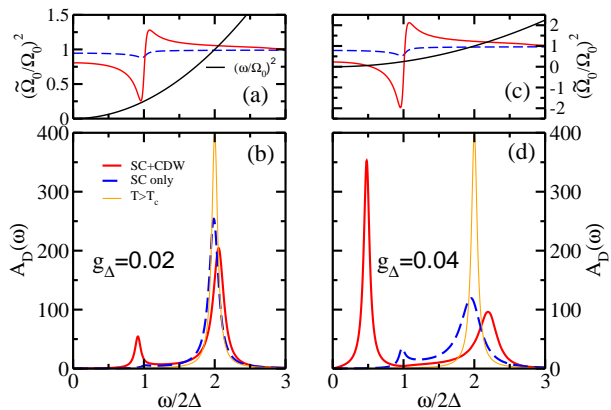}
\caption{Renormalization of the phonon frequency (a,c) and of the phonon
  spectral function $A_D$ (b,d) due to the Higgs mode according to the
  phenomenological approach by LV,\cite{varma_prb82} where the phonon
  self-energy is $\Sigma_\D=-2g^2_\D\chi_{\D\D}/UX_{\D\D}$. The Higgs
  fluctuations $X_{\D\D}$ are computed in the two cases, with (solid red
  lines) or without (dashed blue lines, as done in Ref.\
  [\onlinecite{varma_prb82}]) a preformed CDW, and correspond to the data
  of Fig.\ \ref{fig-prop} for $\g=0.1\D_0$. Here the CDW phonon has energy
  $\O_0=4\D_0$ and $\d=0.1\D_0$, leading to the spectral function at
  $T>T_c$ shown by the thin orange line in panels (b) and (d). When
  entering the SC phase the phonon peak at $\O_0$ gets broader and slightly
  displaced, while a second peak develops below $2\D_0$ due to the Higgs
  self-energy $\Sigma_\D$, that causes a second intersection for the curve
  $\o^2-\tilde\O_0^2$ in the inverse phonon propagator.}
\label{fig-varma}
\end{figure}

Below $T_c$ the opening of the SC gap modifies both the Raman response,
that now inlcudes also the contribution \pref{chifull} of the Higgs mode,
and the phonon propagator itself, that gets renormalized by the SC
amplitude fluctuations.\cite{varma_prb82,brouwne_prb83} The latter
mechanism was proposed originally by Littlewood and Varma
(LV),\cite{varma_prb82} who introduced a phenomenological coupling $g_\D$
bewteen the Higgs and the phonon.  By following the language of LV, one can
then write the phonon propagator $D$ in the mixed state as
$D^{-1}(\O_n)=-\frac{\O_n^2+\tilde\Omega_0^2}{2\Omega_0}$, where
$\tilde\O_0^2=\O_0^2[1-2\Sigma_\Delta/\O_0]$ is the phonon energy
renormalized by the coupling of the phonon to Higgs fluctuations, so that
$\Sigma_\D(\o)= -2g_\D^2 {\chi_{\D\D}(\o)}/{UX_{\D\D}}$. Since near $2\D_0$
one has $\chi'_{\D\D}\simeq -2/U$ while $X'_{\D\D}\ra 0$, as shown in Fig.\
\ref{fig-prop}, the pole of the phonon propagator determined as usual by
$\o^2-\tilde\O^2(\o)=0$ has a new solution around the frequency of the
Higgs mode, see Fig.\ \ref{fig-varma}a,c. Once more the nature of the
Higgs, encoded in $X'_{\D\D}$ and $\G_{\D\D}$, gives qualitative and
quantitative differences if one applies the above phenomenological approach
by using the standard form of the Higgs fluctuations, as done by LV, or the
real one in the coexisting SC+CDW state. These differences are elucidated
in Fig.\ \ref{fig-varma}, where we show the renormalized phonon frequency
$\tilde\O_0$ and the phonon spectral function for two values of the
phenomenological coupling $g_\D$ and a residual phonon broadening
$\d=0.1\D_0$ above $T_c$.  Here the Higgs fluctuations in the two cases (SC
and SC+CDW) correspond to the calculations shown in Fig. \ref{fig-prop} for
$\g=0.1\D_0$. As one can see, for the same remaining parameters a weak
feature found for conventional Higgs fluctuations turns out in a strong
feature when the Higgs is computed in the mixed state.\cite{nota_varma} Notice that the
crucial role of the residual damping $\g$ has been neglected so
far,\cite{varma_prb82,brouwne_prb83} while it is certainly present in real
materials and suppresses the signature of a conventional Higgs mode even
when the coupling to the phonon moves it inside the quasiparticle
continuum. Thus, already this phenomenological approach shows that it is
crucial to retain the real nature of the Higgs mode in the coexisting state
in order to explain the strong signatures observed
experimentally.\cite{sacuto_prb14}

As shown later on by Browne and Levin (BL),  the effective coupling $g_\D$ between the
phonon and the Higgs arises microscopically by the coupling between
the amplitude fluctuations of both the CDW and SC order parameter, so
that the phonon propagator below $T_c$ reads:
\be
\lb{om0new}
D^{-1}(i\O_n)=-\frac{\O_0^2-(i\O_n)^2-2g^2\o_0\chi_{D\Delta}^2/X_{\D\D}}{2\o_0}
\ee
where the function
\be
\lb{chirddelta}
\chi_{ D\D}=-{8 D_0\D_0}\sum_\bk
\frac{\g_\bk^2}
{E_\bk((i\O_n)^2-4E_\bk^2)}\tanh(\beta
E_\bk/2)
\ee
is the one that mediates the effective coupling between the phonon and
the Higgs fluctuations. It is worth noting that in their paper, BL use a model
where the SC and CDW order parameters coexist only on some fraction
$\eta$ ($\g$ in the notation of Ref.\ [\onlinecite{brouwne_prb83}]) of
the Fermi surface. In the limit when $\eta\ra 0$ BL notice that their
result reproduces the one by LV: indeed, in this limit the Higgs mode
is the standard one, since in Eq.\ \pref{xddapp} above there is no CDW
gap above the SC one to push the quasiparticle continuum away from the
Higgs pole. When instead $\eta$ increases BL observe some numerical
difference with respect to the results of LV, that they uncorrectly
attribute to a larger value of the effective coupling to the phonon,
i.e. the function $\chi_{D\D}$ in Eq.\ \pref{chirddelta}. However,
this is not the case: indeed, even if a larger overlap bewteen the SC
and CDW order parameters leads to an increase of $\chi_{D\D}$, the
stronger effect in the formation of the coexistence state is in the
profound modification of the Higgs spectral function. This is clearly
shown in Fig.\ \ref{fig-varma}, where the two cases (Higgs mode
computed with or without the CDW gap) are compared by keeping the same
effective coupling $g_\D$ of the phonon to the Higgs.

\begin{figure}[htb]
%full.eps
\includegraphics[width=8.5cm,clip=]{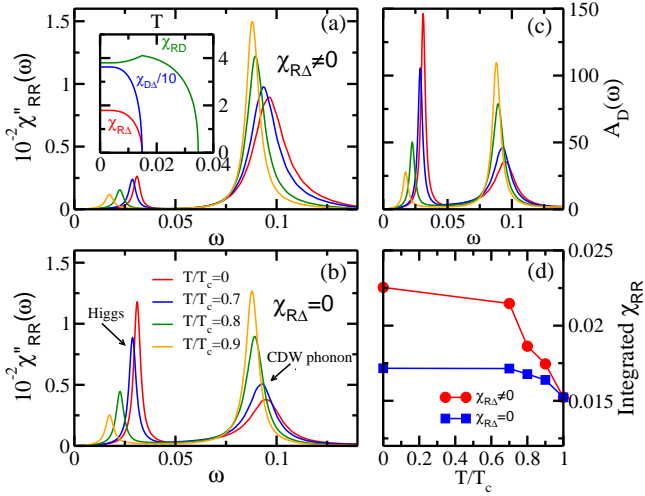}
\caption{Raman response in the $A_{1g}$  channel in the mixed CDW+SC state
  according to Eq.\ \pref{raman_full}, both in the presence (a) and
  in the absence (b) of the direct Raman coupling to the Higgs, 
  encoded in $\chi_{R\D}$. The double-peak structure found below $T_c$ in
  the phonon spectral function, shown in panel (c), is weighted in the
  Raman response by the two polarization functions $\chi_{R\D}$ and
  $\chi_{RD}$. Their value at $\o=0$ is shown in the inset of
  panel (a), along with the coupling $\chi_{D\D}$ between the two order
  parameters responsible for the Higgs siganture in the phonon mode.
 (d) Temperature evolution of the Raman spectral weight integrated between
$\o=0$ and $\o=0.15$, for the two cases of panels (a)-(b).} 
\label{fig-full}
\end{figure}

Once clarified the role of the modified Higgs fluctuations on the phonon
spectral function we now present the results for the Raman response in
the coexisting state. As one can easily see, the coupling \pref{chirddelta} between the two order parameters
renormalizes also the intermediate process $\chi_{RD}$ making the
phonon Raman visible, so that the full Raman response reads:
\bea
\lb{raman_full}
\chi_{RR}&=&-\frac{\chi^2_{R\D}}{X_{\D\D}}-\frac{2g^2\o_0\left[\chi_{RD}-\frac{\chi_{R\D}\chi_{\D
    D}}{X_{\D\D}}\right]^2}
{\O_0^2-(i\O_n)^2-2g^2\o_0\chi_{D\Delta}^2/X_{\D\D}},
\eea
where the first term corresponds to the response \pref{chifull} of the
Higgs alone and the second one accounts for the response \pref{rrphonon} of
the phonon mode, renormalized by the Higgs fluctuations according to Eq.\
\pref{om0new}. The temperature evolution of the Raman spectra is shown in
Fig.\ \ref{fig-full}a for the $\D_0$, $D_0$ appearing in Fig.\
\ref{fig-prop}-\ref{fig-scd}.  In agreement with the discussion above, the
phonon spectral function, shown in Fig.\ \ref{fig-full}c, shows a
double-peak structure evolving with temperature, one around $\O_0$ and one
below $2\D_0$, that is the signature of the Higgs mode. However, the way
these two peaks appear in the Raman response depends crucially on the
polarization functions $\chi_{RD}$ and $\chi_{R\D}$, see Fig.\
\ref{fig-full}a,b. In particular, by comparing the results of Fig.\
\ref{fig-full}a with the ones in panel b, when we put by hand
$\chi_{R\D}=0$, one sees that the direct coupling $\chi_{R\Delta}$ of the
Higgs to the light is crucial to establish the Raman spectral-weight
distribution below $T_c$. Indeed, when $\chi_{R\D}=0$ the Higgs feature at
$T=0$ is stronger than in the case of $\chi_{R\D}\neq 0$, shown in Fig.\
\ref{fig-full}a, and the broadening of the phonon peak is also
smaller. Both effects are due to the vanishing of the second term in square
brackets of Eq.\ \pref{raman_full}, that partly screens out the response at
the Higgs when present. Nonetheless, the increase of the Raman spectral
weight integrated up to $\o\simeq 0.15$, shown in Fig.\ \ref{fig-full}d, is
larger when $\chi_{R\D}\neq 0$, due to a larger broadening of the phonon in
this case.  Notice also that from Eq.\ \pref{raman_full} one immediately sees 
that the peak at the Higgs due to the first term of Eq.\ \pref{raman_full},
and represented in Fig.\ \ref{fig-scd} for the case $g=0$, cancels out here
with the second term. Thus in this case the Higgs becomes visible only
trough its coupling to the phonon, but its direct visibility $\chi_{R\D}$
still influences the overall shape of the Raman spectra.

The crucial difference between the behavior of the Raman response and
the behavior of the phonon propagator in a phonon-induced CDW
transition has been completely overlooked in the previous
work.\cite{varma_prb82,brouwne_prb83} For example, the claim usually
done\cite{varma_prb82,brouwne_prb83} that the transfer of spectral
weight bewteen the two peaks in the phonon spectral function $A_D$
must be found also in the Raman spectra is not correct. In general,
this is not the case, as the comparison bewteen the panels a,b,c of
Fig.\ \ref{fig-full} clearly demonstrates. This simple fact can also
be used to explain the polarization dependence of the Higgs signatures
reported recently in NbSe$_2$.\cite{sacuto_prb14} Indeed, even though
our model is not intended to give a realistic description of NbSe$_2$,
nonetheless some general conclusions can be drawn from our results
that can be used to interpret the Raman experiments. The general results of
our calculations in the case of a phonon-mediated Higgs response can
be summarized in three separate items. First, the overlap bewteen the
CDW and SC order parameter in part of the Fermi surface is necessary
to give rise to a finite $\chi_{D\D}$ function in Eq.\ \pref{chirddelta}, that controls the
effective coupling between the CDW phonon and the Higgs, see Eq.\ \pref{om0new}. In turn,
larger is the fraction of Fermi surface where the overlap occurs and
stronger is the enhancement of the Higgs mode itself, shown in Fig.\
\ref{fig-prop}, that helps its visibility. Second, both the phonon and the
Higgs become Raman visible trough an intermediate electronic
processes, given by $\chi_{RD}$ and $\chi_{R\D}$
respectively, see Fig.\ \ref{fig-diagrams}. These two functions depend on the combined symmetry of
the (CDW or SC) gap and of the Raman light polarization. Thus,
different Raman symmetry will give in general a different result,
since both functions will acquire a specific frequency and temperature
dependence. As a consequence, the differences observed experimentally in NbSe$_2$
by means of the various Raman symmetries\cite{sacuto_prb14}
can be used to extract specific informations on the response functions
$\chi_{RD}$ and $\chi_{R\D}$, and in turn on the gaps themselves.

\section{Discussion and Conclusions}

The present work focuses on two different problems: the nature of the
Higgs mode in the coexisting CDW+SC state, and the issue of its
detection in Raman spectroscopy. The first result, discussed in
Sec. II, is that in the coexisting CDW+SC state the CDW gap pushes the
quasiparticle excitations away from the Higgs pole. Thus, in contrast
to the usual SC case, the Higgs mode presents a sharp spectral
function, whose detection can be easier, whatever is the mechanism
that couples it to a physical observable. It is also interesting that
the power-law divergence of the amplitude fluctuations at twice the SC
gap value resembles the behavior of the Higgs mode predicted in
relativistic ${\cal O}(N)$
theories,\cite{podolsky_prb11,prokofev_prl12,podolsky_prl13} that are
expected to work for superfluids or strongly-coupled
superconductors. A possible way to test directly our prediction could
be for example an out-of-equilibrium optical experiment, where a
non-linear coupling of the current to the Higgs can be
generated.\cite{shimano14} Indeed, in this case we expect that the
strongly reduced damping of the Higgs mode in the coexisting state
will lead to a much slower decay of the amplitude oscillations with
respect to the conventional superconductors investigated so
far.\cite{shimano,shimano14} It is also worth noting that the
mechanism outlined here can also have a potential impact on the
detection of the Higgs mode in other families of superconductors. For
example in cuprate superconductors, where
signatures of a $2\D_0$ oscillations in out-of-equilibrium
spectroscopy have been also
reported, \cite{carbone} the pseudogap in the quasiparticle
excitations present already above $T_c$ could
contribute to enhance the Higgs fluctuations. Indeed, as mentioned
above, a model similar
to the one studied in the present paper has been studied long ago as a
simplified phenomenological description of a CDW-like pseudogap
phase.\cite{nota_ddw} The quantitative effect on the Higgs 
should be however estimated within realistic models, since in these
materials the quasiparticle continuum extends up to zero frequency
due to the $d$-wave nature of the order parameter. Finally, it can be
worth exploring also the nature of the Higgs fluctuations in the
coexisting spin-density-wave and superconducting state, a situation
realized nowadays is some families of iron-based
superconductors.\cite{review_pnictides} 

The second part of the paper focuses on the Raman detection of the Higgs
mode.  Here the final results depend crucially on the nature of the CDW
instability itself, as due to an electronic interaction or to the coupling
to a phonon. In the former case one probes directly the Higgs spectral
function, so a signature at $2\D_0$ appears in the Raman response, while in
the latter case the Higgs signature moves below $2\D_0$ due to the coupling
to the phonon. In both cases one sees that when damping effects, neglected
in the previous literature,\cite{varma_prb82,brouwne_prb83} are taken into
account, it is crucial to consider the modified nature of the Higgs mode
in the coexisting state in order to resolve the Higgs signatures in the
experiments.  Moreover, we clarified how the Raman response depends
crucially on the intermediate electronic processes that make the Higgs or
the phonon Raman visible. This issue, that is often overlooked in the
literature,\cite{varma_prb82,brouwne_prb83,varma_cm14}  explains the origin of the direct Raman visibilty of the Higss
and the transfer of spectral weight between the phonon and Higgs peak. In
particular, we suggest that our results can be used as a guideline to infer useful
information on the still unsolved issue of the symmetry of the CDW and SC
gaps in NbSe$_2$.

\section{Acknowledgements}
We thank M. M\'easson, C. Castellani and J. Lorenzana
for useful discussions and suggestions. We acknowledge financial support by MIUR under the
projects FIRB-HybridNanoDev-RBFR1236VV, 
PRIN-RIDEIRON-2012X3YFZ2  and Progetto Premiale 2012-ABNANOTECH.

\appendix

\section{Coupling of the Higgs mode to Raman in a lattice model}

In this Appendix we clarify the role played by the intermediate
electronic process $\chi_{R\D}$ that couples the Higgs mode to the
Raman charge density, see Fig.\ \ref{fig-diagrams}a. In particular, we show that at weak SC coupling
the coupling of the SC amplitude fluctuations to the Raman charge
fluctuations ($\chi_{R\D}$) or to the real charge fluctuations
($\chi_{\rho\D}$) can be different from each other. Such a difference
is crucial since these two quantities play, as we shall see, a quite
different role.

Let us first discuss the coupling of the Higgs mode to real charge
fluctuations in an ordinary SC, i.e. without preformed CDW state.
By means of the standard definition of charge operator $\rho(\bq)\equiv \sum_{\bk\s} 
c^\dagger_{\bk-\bq/2,\s}c_{\bk+\bq/2,\s}$, one can easily find the
well known\cite{randeria,benfatto_prb04} result that:
\be
\lb{chirdsc}
\chi_{\rho\D}=\frac{4\D_0}{N}\sum_\bk
\frac{\xi_\bk \tanh(\b E_\bk/2)}{E_\bk[(i\O_n)^2-4E_\bk^2]}.
\ee
where $\xi_\bk=\e_\bk-\mu$ is a generic band dispersion and
$E_\bk=\sqrt{\xi_\bk^2+\D_0^2}$. In the weak-coupling limit the above
integral is dominated by the region around the Fermi level,
$\xi_\bk=0$. As a consequence, for {\em any} band structure one can
assume that the above integral vanishes because of the particle-hole
symmetry of the integration limits enforced by the BCS solution, that
allows one to put $\chi_{\rho\D}\simeq N_0\D_0\int_{-\infty}^\infty
d\xi \, \xi/[\sqrt{\xi^2+\D_0^2}(\o^2-4 \xi^2-4\D_0^2)]= 0$. In general, the
particle-hole symmetry of the fermionic bubbles in the BCS limit is
also evocated to guarantee that amplitude and charge/phase
fluctuations are uncoupled at Gaussian level, as we mentioned above
Eq.\ \pref{ssc}, allowing one to study the amplitude sector
independently from the charge one. The situation is instead radically different at
strong coupling, as discussed e.g. in Ref.\
[\onlinecite{randeria,benfatto_prb04}]. 

When one computes instead the coupling of the Higgs mode to the Raman
charge fluctuations $\rho_R(\bq)\equiv \sum_{\bk\s} \G(\bk)
c^\dagger_{\bk-\bq/2,\s}c_{\bk+\bq/2,\s}$ one should consider the additional momentum
modulation of the charge provided by the polarization dependent factor
$\Gamma(\bk)$. Thus, even remaining within a BCS scheme
$\chi_{R\D}$ can be different from zero even if $\chi_{\rho\D}\simeq
0$. This can be seen easily in our model, where the two quantities
$\chi_{R\D}$ and $\chi_{\rho\D}$ at half filling are given
respectively by Eq.\ \pref{chird} and:
\be
\lb{chirdelta}
\chi_{\rho\D}=\frac{4\D_0}{N}\sum_\bk
\frac{\e_\bk \tanh(\b E_\bk/2)}{E_\bk[(i\O_n)^2-4E_\bk^2]},
\ee
where now $E_\bk=\sqrt{\e_\bk^2+\D_0^2+D_0\g_\bk^2}$. As a
consequence in our case,  where the band structure is particle-hole symmetric, at half
filling $\chi_{\rho\D}$ is {\em exactly zero} and it reamains
negligibly small away from it. On the other hand, as
we mentioned in the text, in the $A_{1g}$ channel where $\G(\bk)=(\cos
k_x+\cos k_y)\propto \e_\bk$ the integral \pref{chird} that defines
$\chi_{R\D}$ is finite.  Indeed, for a
constant CDW gap,  Eq.\ \pref{chirdelta} at $T=0$ can be approximately
estimated as $\chi_{\rho\D}\sim \D_0N_0 \int d\e \, \e
/[\sqrt{\e^2+R^2}(\o^2-4\e^2-4R^2)]=0$ while 
$\chi_{R\D}\sim \D_0N_0\int d\e \,
\e^2/[\sqrt{\e^2+R^2}(\o^2-4\e^2-4R^2)]\neq 0$, where
$R^2=\D_0^2+D_0^2$. Notice that to appreciate this difference it is
crucial to retain a full lattice description of the problem, that has
been neglected in the previous theoretical approach.\cite{varma_prb82,brouwne_prb83}

\begin{figure}[htb]
%chirdelta.eps
\includegraphics[width=7cm,clip=true]{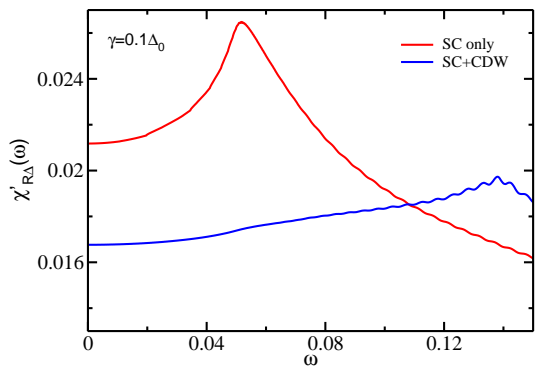}
\caption{Real part of the Raman response function $\chi_{R\D}$ at $T=0$ in the pure
  SC and in the coexistence SC+CDW state for a residual $\gamma=0.1\D_0$,
  as in Fig. 1 and Fig. 2. As one can see, in the mixed CDW+SC case
  $\chi'_{R\D}$, that appears as a prefactor in the Raman susceptibility
  \pref{chirrapp}, is featureless at the frequency $2\Delta_0$ where the
  Higgs spectral function shown in Fig.\ \ref{fig-prop}c has a maximum.} 
\label{fig-chird}
\end{figure}

In summary, in our description of the CDW+SC
state we can still safetly neglect all the complications\cite{randeria,benfatto_prb04} of the coupling between the amplitude
mode and the charge/density one, since we are still in a weak-coupling
scheme where $\chi_{\rho\D}\simeq 0$ at all dopings, but we can retain
a finite coupling of the Higgs to the Raman probe, encoded in
$\chi_{R\D}\neq 0$. In this respect, the general claim done in Ref.\
[\onlinecite{varma_cm14}] that the Higgs
mode is decoupled by the Raman spectra because of particle-hole
symmetry is formally uncorrect. However, it is still true that
$\chi_{R\D}$ is a small quantity, so when the Higgs is an overdamped
mode, as in an ordinary superconductor, the overall effect of amplitude fluctuations
on the Raman response will be negligible, see dashed lines in Fig.\ \ref{fig-scd}. In the CDW+SC state instead
the modifications of the Higgs spectral function $A_\D$ can make this small coupling
$\chi_{R\D}$ crucial, leading to the strong Raman response represented
by the solid lines in Fig.\ \ref{fig-scd}. We notice also that the
enhancement of the Raman response in the mixed state occurs here
despite the fact that $\chi_{R\D}$ itself is {\em smaller} in the
coexisting state, as shown in Fig.\ \ref{fig-chird}, due to the
presence of the CDW gap. 
In other words, as we emphasized above, the crucial role of the CDW
for what concerns the Raman response is {\em not} to enhance the visibility itself of the
Higgs, encoded in the electronic process $\chi_{R\D}$ (that is indeed
even suppressed). The CDW changes crucially the nature of the Higgs
spectral function, making the Higgs detectable even in the presence of
a small coupling to the Raman light.

\section{Variation of the Higgs mode with doping}

\begin{figure}[htb]
%phase-diagram.eps
\includegraphics[width=8.5cm,clip=]{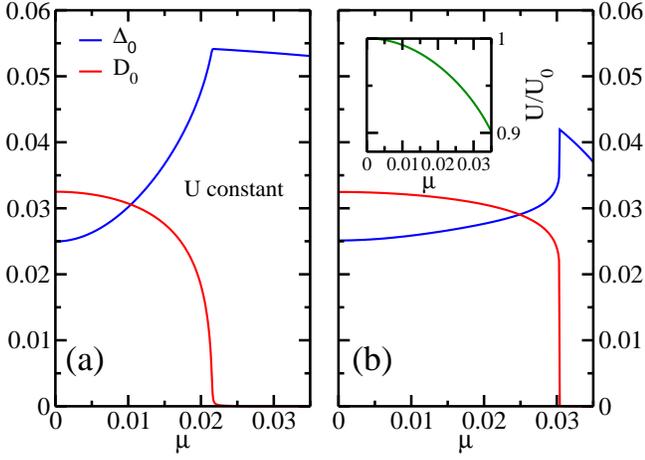}
\caption{Doping variation of the CDW and SC gap at $T=0$ in the case of a constant (a) or varying (b) SC coupling $U$. In (b) $U$ decreases slightly with doping with respect to the value $U_0$ at $\mu=0$ (see inset), in order to reproduce a weaker increase of $\D$ with doping.} 
\label{fig-delta}
\end{figure}

To elucidate the effect of the CDW gap on the Higgs mode in the case
where a larger part of the Fermi surface remains ungapped below
$T_{CDW}$ we show here the analogous calculations of Fig. 2 of the
manuscript away from half-filling. In our model the CDW gap decreases
when one moves away from half-filling, due to the lack of perfect
nesting. If one keeps the SC coupling $U$ fixed this reflects also in
a rapid increase of the SC order parameter at $T=0$, as shown in Fig.\
\ref{fig-delta}a. However, in order to make the comparison between
different dopings meaningful, we decided here to reduce also $U$ with
doping, in order to retain an almost constant SC gap, see Fig.\
\ref{fig-delta}b. This allows one to compare the effects on the Higgs mode
due only to the increase of the quasiparticle contribution starting at
$2\D_0$, while keeping $\D_0$ almost fixed. 

As mentioned in Sec. II, when one moves away from
half-filling the general form of the effective action is more
complicated than Eq. \pref{sapprox} of the manuscript, since one should retain in
principle also the coupling of the amplitude mode at $\bq=0$ and
$\bq=\bQ$. One can then show that the Raman response reads:
\be
\lb{full}
\chi_{RR}(q)=\left[ \mathcal{M}^0_{RR}(q)-\mathcal{M}_{\Delta R}(q)\mathcal{M}^{-1}_{\Delta \D}
\mathcal{M}^T_{\Delta R}(-q)\right]_{11}
\ee
where the matrices $\mathcal{M}^0_{RR}$, $\mathcal{M}_{\Delta R}$ and $\mathcal{M}_{\Delta \D}$ correspond to the BCS term, to the Higgs-Raman coupling and to the Higgs fluctuations, respectively, with:  
\begin{equation}\label{DELTA_J_MATRICES}
	\mathcal{M}^0_{R R}(q)\equiv
		\begin{bmatrix}
		\Lambda^{\Gamma\Gamma}_{3333}(q)&&\Lambda^{\tilde{\Gamma}\Gamma}_{1333}(q)\\\\
		\Lambda^{\Gamma\tilde{\Gamma}}_{3313}(q)&&\Lambda^{\tilde{\Gamma}\tilde{\Gamma}}_{1313}(q)
		\end{bmatrix},
\ee
\be
\mathcal{M}_{\Delta R}(q)\equiv
\begin{bmatrix}
\Lambda^\Gamma_{0133}(q)&&\Lambda^\Gamma_{1133}(q)\\\\
\Lambda^{\tilde{\Gamma}}_{0113}(q)&&\Lambda^{\tilde{\Gamma}}_{1113}(q)
\end{bmatrix},
\end{equation}
\begin{equation}
		\mathcal{M}_{\Delta\Delta}(q)\equiv
		\begin{bmatrix}
		\frac{2}{U}+\Lambda_{0101}(q)&&\Lambda_{1101}(q)\\\\
		\Lambda_{0111}(q)&&\frac{2}{U}+\Lambda_{1111}(q)
		\end{bmatrix}.
\end{equation}
The fermionic  susceptibilities $\Lambda_{ijkl}(q)$ are defined as
\begin{equation}
	\Lambda_{ijkl}(q)\equiv \frac{T}{N}\sum_k\text{Tr}\left[G_0(k+q)\sigma_i\otimes\sigma_j
	G_0(k)\sigma_k\otimes\sigma_l\right],
	\end{equation}
and the superscripts $\G,\tilde \G$ refer to the insertion of a Raman vertex $\G(\bk)$ or $\tilde\G(\bk)\equiv\G(\bk+\bQ/2)$. 

\begin{figure}[htb]
%doping.eps
\includegraphics[width=8.5cm,clip=]{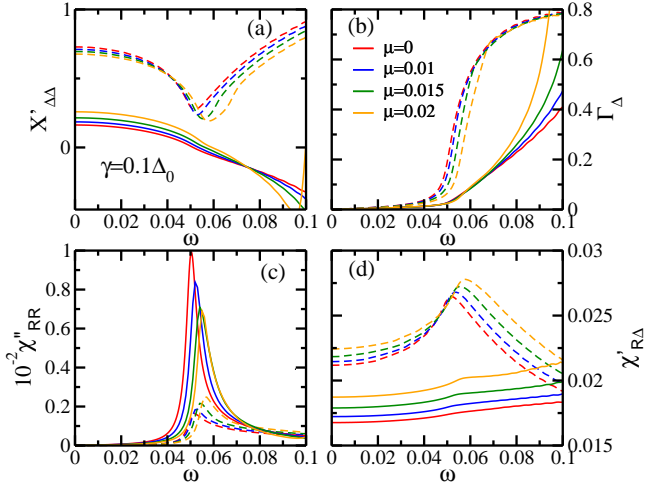}
  \caption{Frequency dependence of the real (a) and imaginary (b) part
    of the inverse amplitude mode at $T=0$ for different doping
    levels. In the CDW+SC state the gaps are taken from Fig.\
    S\ref{fig-delta}b, while in the pure SC case $U$ is tuned to have
    the same $\D_0$ at $T=0$. The resulting Raman response is shown in
    panel (c), while panel (d) shows the Raman-Higgs coupling
    function, that is always smaller in the coexistence state.}
\label{fig-filling}
\end{figure}

The element $[\mathcal{M}_{\D\D}]_{11}\equiv X_{\D\D}$ corresponds at
half-filling to the inverse Higgs propagator given by Eq. \pref{xddapp}.
 Moreover, at $\mu=0$ in the $A_{1g}$ channel the
off-diagonal elements of both $\mathcal{M}_{\Delta R}$ and
$\mathcal{M}_{\Delta \D}$ vanish, and one is left with the Eq.\
\pref{chifull}of Sec. II above.  At $\mu\neq 0$ we computed the full expression
\pref{full}, even though we verified that the main contribution of the
Higgs fluctuations to the Raman response comes from $X_{\D\D}$, whose
real and imaginary parts are displayed in Fig.\
\ref{fig-filling}a,b. As one can see, even though at finite filling a
finite Fermi surface exists above $T_c$, leading to a larger
contribution of the quasiparticle continuum above $2\D_0$, still the
Higgs mode preserves a strong relativistic character near the pole,
with a progressively larger damping above it. Thus, when one compares
again the results in the pure SC and in the mixed CDW+SC state, Fig.\
\ref{fig-filling}c, in the latter case the Higgs spectral function
leads to a stronger feature in Raman. In full analogy with the case of 
Fig.\ \ref{fig-scd}, this enhanced Raman response is not due to
the prefactor, represented in the case of Eq.\ \pref{full} by the
matrix $\mathcal{M}_{\D R}$, whose largest contribution
$\chi_{R\D}\equiv [\mathcal{M}_{\D R}]_{11}$ is shown in Fig.\
\ref{fig-filling}d. Once more, the enhancement of the response in the mixed
state reflects the enhancement of the Higgs fluctuations due to the
preformed CDW gap, even when this leaves larger part of the Fermi surface
ungapped. We checked that the present results hold
also when the Coulomb interaction is explicitly taken into account,
since both the BCS term $\mathcal{M}^0_{RR}$ and the screening term
coming from charge fluctuations\cite{review} are always much smaller
than the contribution of the Higgs.  Finally, we note that even though
the present manuscript is not intended to give a quantitative
description of NbSe$_2$, the persistent anomalous character of the
Higgs mode with doping shown in Fig.\ \ref{fig-filling} is an
encouraging suggestion that the same results will hold also with a
more general band structure as the one of NbSe$_2$, where finite
portions of the Fermi surfaces are ungapped by the CDW. In addition,
Fig.\ \ref{fig-delta}b shares also some similarity with the
experimental phase diagram under pressure of
NbSe$_2$\cite{pressure1,pressure2}, suggesting that the present
approach could also be used in the future for a qualitative
understanding of the evolution of the Raman spectra under pressure.


\begin{thebibliography}{cc}
\bibitem{nota_varma}
To compare the results of Fig.\ \ref{fig-varma} with the ones in Ref.\
[\onlinecite{varma_prb82}] consider that the values $g_\D=0.02$ and
$g_\D=0.04$ used here correspond to the values $\a=0.25$ and $\a\simeq
1$ of the dimensionless coupling $\a$ introduced by LV.
\end{thebibliography}
\end{document}